\documentclass[aps,twocolumn,showpacs]{revtex4}
\usepackage{graphicx}

\begin{document}

\title{On The Universal Scaling Relations In Food Webs}

\author{L. A. Barbosa$^{\ddagger}$, A. Castro e Silva$^{\ast}$ and J.
 Kamphorst Leal da Silva$^{\dagger}$}

\affiliation{
Departamento de F\'\i sica,Instituto de Ci\^encias Exatas,
 Universidade Federal de Minas Gerais\\
C. P. 702, 30123-970, Belo Horizonte, MG, Brazil\\}

\date{\today}\widetext


\begin{abstract}
In the last three decades, researchers have tried to establish universal
patterns about the structure of food webs. Recently was proposed that
the exponent $\eta$ characterizing the efficiency of the energy
transportation of the food web had a universal value ($\eta=1.13$).
Here we establish a lower bound and an upper one for this exponent
in a general spanning tree with the number of trophic species and the trophic levels
fixed. When the number of species is large the lower and upper bounds
are equal to $1$, implying that the result $\eta=1.13$  is due to finite
size effects. We also evaluate analytically and numerically the exponent $\eta$ for 
hierarchical and random networks. 
 In all cases the exponent $\eta$ depends on the number of trophic species $K$
and when $K$ is large we have that $\eta\to 1$. Moreover, this result holds for any number $M$
of trophic levels. This means that food webs are very efficient resource transportation
systems.

\end{abstract}

\maketitle

Understanding energy and material fluxes through ecosystems is
central to many questions in ecology~\cite{1,2,3}. Ecological
communities can be studied via resource transfer in food
webs~\cite{4}. These webs are diagrams showing the predation
relationships among species in a community. Usually, a group of
species sharing the same set of predators and the same set of prey
is aggregated in one trophic specie~\cite{5,6}. So, each
trophic specie is represented by a site and denoted by an integer
number $i=1,...~,K$, where $K$ is the total number of trophic
species. A relation between a pair of sites is represented by a
link directed from the prey to predator. There are several
quantities introduced in the literature to characterize the food
web structure, such as the fractions of the species in the trophic
levels (basal, intermediates and top), the fractions of links
among them, the connectance, the average distance between two
sites, the clustering coefficient and the degree distribution. It
turns out that all these quantities are nonuniversal~\cite{Garla2}
and dependent of the size of the food web. Perhaps, the only variable
in common agreement with the literature is the maximum
value of trophic levels $(M \leq 4)$. Garlaschelli et
al.~\cite{Garla1} have considered food web as transportation
networks~\cite{banavar,west} whose function is to deliver
resources from the environment to every specie in the network. In
this case, food webs appear to be very similar to other systems
with analogous function, such as river and vascular networks. In
that work they have represented a real food web by spanning trees
with minimal lengths. For each specie $i$ the number $A_{i}$ of
species feeding directly or indirectly on $i$, plus itself, is
computed. They also computed the cost of this transfer, namely
$C_{i}=\sum_{k}A_{k}$, where $k$ runs over the set of direct an indirect 
predators of $i$ plus itself. 
 The shape of $C_{i}$ as a function of $A_{i}$
follows a power law relation $C\left(A \right) \sim A^{\eta}$,
where the scaling exponent $\eta$ quantifies the degree of
optimization of the transportation network. They found the same 
allometric scaling relation for different food webs. By plotting 
$C_{i}$ versus $A_{i}$ for each one of the seven large food webs 
in the literature, and  by plotting $C_{0}$ versus $A_{0}$ for a 
set of different food webs. The exponent found, varies between  $1.13$
and $1.16$. Therefore, they concluded that the exponent $\eta$ has
a universal value ($\eta=1.13$) and it is, perhaps, the only
universal quantity in food webs. Nevertheless, this matter has
been the subject of debates~\cite{camacho,Garla3}.

Here we establish an upper bound ($\eta_{max}$) and a lower one ($\eta_{min}$)
for the exponent $\eta$ in a general spanning tree with $M$ trophic levels and $K$
trophic species, both fixed. In the limit $K\to\infty$, we have that 
$\eta_{max}=\eta_{min}\to 1$.
 We also evaluate analytically and numerically the exponent $\eta$ for hierarchical and
random networks. Our main conclusions are that (a) the result
$\eta=1.13$ for food webs is due to finite size effect (small $K$), (b) the exponent
$\eta$ depends on the $K$ and when $K$ is large we have that
$\eta=1$. Moreover, this results hold for any number $M$ of trophic levels, implying
 that food webs are efficient resource transportation networks.

It is worth mentioning that this problem is related to river and vascular
networks~\cite{banavar}. Consider  $K$ sites uniformly
distributed in a $d$-dimensional volume. The network is constructed by
linking the sites, in such way that there is at least a path
connecting each site to the source (a central site). Since each
site is feed at steady rate $F_i=F$,the metabolic rate $B$
clearly is given by $B=\sum_i F_i=FK$. Let $I_b$ represent the
magnitude of flow on the $b^{th}$ link. Then, the total quantity of
nutrients in the network, at a particular time, is given by
$V=\sum_b I_b$. They define the most efficient class of network as
that for which $V$ is small as possible. Using this procedure they
found that $V \sim B^{(d+1/d)}$. For river basins, $d=2$ and $V
\sim B^{3/2}$. In vascular systems $V \sim B^{4/3}$ since $d=3$.
The variables $A_0$ and $C_0$ of the food webs are related,
respectively, to the number of transfer sites $N$ and to the total
volume of nutrients $V$ by the following equations: $N=A_0-1$ and
$V=C_0-A_0$. Then we have that $C_0 \approx A_0^{(d+1/d)}$ if
$A_0$ is large enough. The value of the exponent $\eta$ for a food
web  can be smaller than the one of rivers ($\eta=3/2$)
or the one of vascular systems ($\eta=4/3$) because the spanning
tree of a food web is not embedded in an Euclidean space.

Let us consider a hierarchical network with $M$ trophic levels.
The network is constructed in the following way. We begin with a
site representing the environment, the site $0$. Then we connect $n_{1}$ sites
to it, since these sites are feeding directly of the environment
they constitute the first trophic level. Obviously, the number of
species in this level is $N_{1}=n_{1}$. The second level is
constructed by connecting $n_{2}$ sites to each site of the first
level. Now, in this level, we have $N_{2}=n_{1}n_{2}$ species.
This procedure is repeated until the level $M$.

Since $A_{i}$ is the number of species feeding directly or
indirectly on site $i$, plus itself, we have that

\begin{eqnarray*}
A_{M}&=&1\\
A_{M-1}&=&n_{M}A_{M}+1=n_{M}+1\\
A_{M-2}&=&n_{M-1}A_{M-1}+1=n_{M}n_{M-1}+n_{M-1}+1\\
\vdots&&\\
A_{0}&=&1+\sum^{M}_{\alpha=1}N_{\alpha}=K+1~~.\\
\end{eqnarray*}

The cost of resource transfer, defined by $C_{i}=\sum_{k}A_{k}$,
where $k$ runs over the set of direct and indirect predators of
$i$ plus itself is given by

\begin{eqnarray*}
C_{M}&=&1\\
C_{M-1}&=&n_{M}C_{M}+A_{M-1}=2n_{M}+1\\
C_{M-2}&=&n_{M-1}C_{M-1}+A_{M-2}=3n_{M}n_{M-1}+2n_{M-1}+1\\
\vdots&&\\
C_{0}&=&1+\sum^{M}_{\alpha=1}(\alpha+1)\prod^{\alpha}_{i=1}n_{i}=1+\sum^{M}_{\beta=1}(1+\beta)N_{\beta}~~.\\
\end{eqnarray*}

\begin{figure}[hbt]
\begin{center}
\resizebox{8,0cm}{5,5cm}{\includegraphics{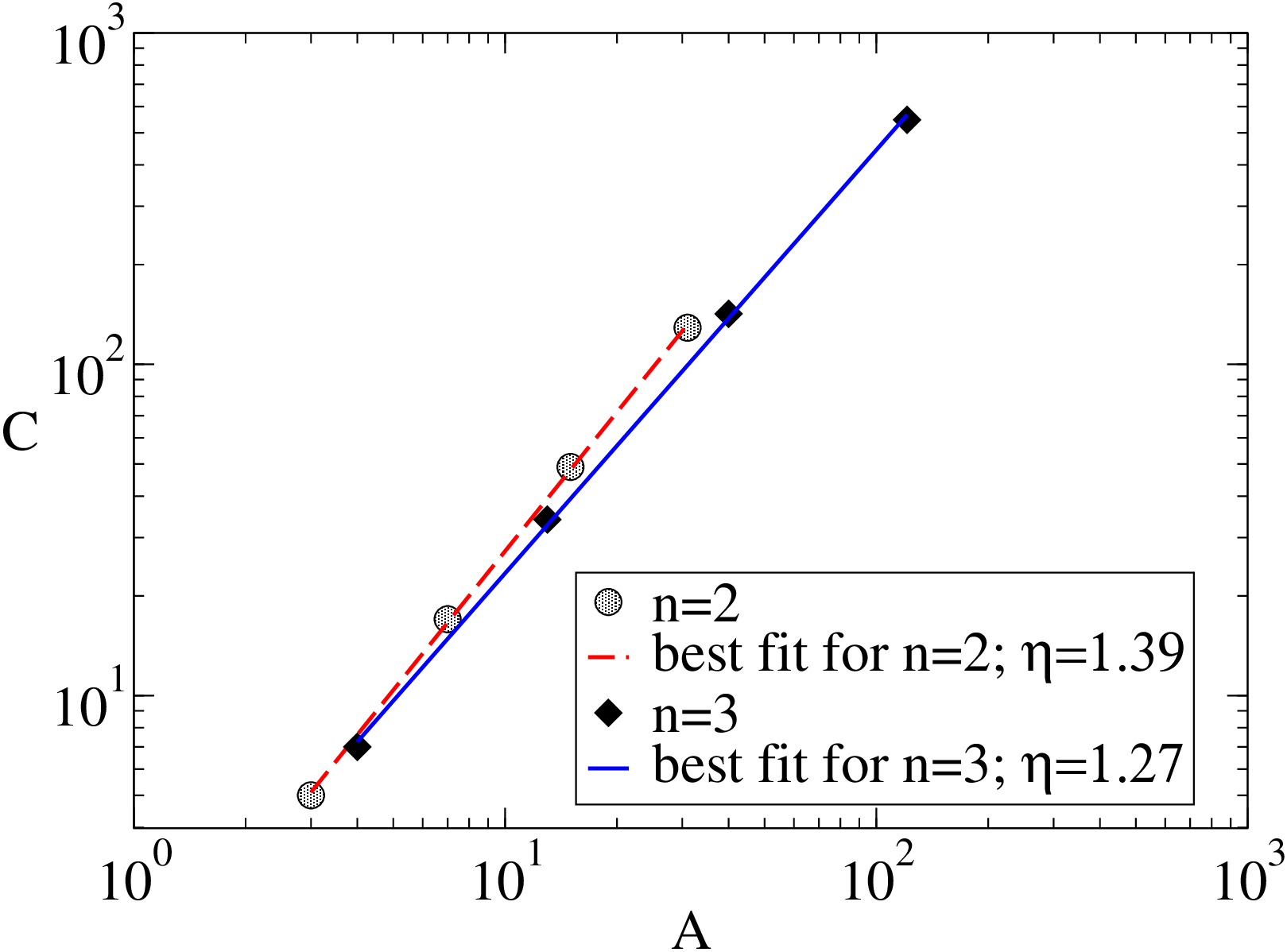}}
\end{center}
\caption{Log-log plots of $C_{i}$ versus $A_{i}$ for networks with constant
ramification ratio $n=2$ and 
$n=3$. Note that the exponent decreases when $n$ grows.}
\label{ration}
\end{figure}

The exponent $\eta$, as was proposed in the
literature~\cite{Garla1}, can be found by (a) plotting $C_{i}$ as
a function of $A_{i}$ for a network with number of trophic level $M$ and
total specie number $K$ fixed; Usually, the point $(1,1)$ is
neglected due to finite size effects.  It can be also found by (b)
plotting $C_{0}$ as a function of $A_{0}$ for several networks
with different trophic species number $K$. This last procedure determines 
the large scale exponent~\cite{Garla3}.
 Note that in hierarchical spanning tree networks, 
$C_{i}$ and $A_{i}$ for species in the same trophic level are equal, implying 
that we have only $M+1$ points in a $C_i\times A_i$ plot.
 Let us firstly use procedure (a) for networks with constant
ramification ratio $n_{i}=n$ and constant number of trophic
levels $M=4$. We find $\eta=1.39$ for $n=2$ and $K=30$ and 
$\eta=1.27$ for $n=3$ and $K=120$, as it is shown in figure ~\ref{ration}.  
 Clearly, the exponent $\eta$ depends on value of $K$, and decreases as 
long as $K$ grows. In the limit that $n\to\infty$, the total number of 
species $K$ also is unlimited and we obtain that the exponent $\eta$ 
approaches the value $1$.

Let us return to the more general case of hierarchical models. The
large scale exponent $\eta$ can be evaluated by,

\begin {equation}
\eta = \frac{\ln C_{0}}{\ln A_{0}}
=\frac{\ln [1+\sum_{\alpha=1}^M (\alpha+1) \prod_{i=1}^\alpha n_i]}
{\ln (1+\sum^{M}_{\alpha=1}\prod_{i=1}^\alpha n_i)}~~.
\label{eq1}
\end{equation}

If at least a ramification ratio is large, $n_\beta\to\infty$,
we have that $\ln A_0\approx\ln n_\beta$ and $\ln C_0\approx \ln n_\beta$. 
 Therefore we find $\eta\to 1$ when the number of species is large.
 We can also use Eq.~\ref{eq1} to evaluate the exponent $\eta$ for
hierarchical networks with constant ramification ration. We find
for this networks $\eta=1.41$ ($n=2$ and $K=30$) and $\eta=1.31$
($n=3$ and $K=120$). These values can be compared with the ones
obtained previously with procedure (a) (see Fig.~\ref{ration}).

In the Eq.~\ref{eq1} the exponent $\eta$ depends on value of $K$,
decreasing as long as $K$ grows. For example, consider the
hypothetical food web with total specie trophic number $K=146$ and
the specie trophic numbers in each level given by $N_1=38$, $N_2=63$,
$N_3=43$ and $N_4=2$. We find the exponent $\eta=1.22$. But, if we
double the number of trophic species in each trophic level
$N_{i}=2N_{i}$ the exponent is now $\eta \approx 1.19$. In that
equation the exponent $\eta$ also depends in the relative
distribution of the species in each level, for a given total
specie number $K$. For the hypothetical food web described above
with $146$ trophic species we change the distributions of species
in each level to $N_1=114$, $N_2=20$, $N_3=10$ and $N_4=2$. We
find the exponent $\eta=1.16$. The exponent has changed from
$\eta=1.22$ to $\eta=1.16$

Now, let us consider a random network with $M$ trophic levels and $K$
trophic species. The network is constructed in the following way.
First, we determine randomly the population in each level $N_\alpha$
($\alpha=1,~2,\ldots,~M$), obeying the restrictions $M$ fixed and
$K$ fixed. Then, the $N_1$ sites are connected to the environment,
constituting the first trophic level. The second level is
constructed by randomly connecting the $N_2$ sites to the $N_1$
sites of the first level. This procedure is repeated until the
level $M$ is constructed. In this case, we can evaluate the mean value 
of $A_i$ and $C_i$ in each level, namely 

\begin{eqnarray*}
\overline{A}_\alpha=\frac{1}{N_\alpha}\sum_{j\in \alpha}A_j
\end{eqnarray*}

\begin{eqnarray*}
\overline{C}_\alpha=\frac{1}{N_\alpha}\sum_{j\in \alpha}C_j~~.
\end{eqnarray*}

Here $\alpha$ specify the trophic level ($\alpha=1,\ldots,~M$).
These quantities are averaged on several random configurations.
Note that in the last level we have that 
$\overline{A}_M=A_i=C_i=\overline{C}_M=1$ and that we always
neglect the point $(1,1)$ in all fits

\begin{figure}[hd]
\centerline{\includegraphics[width=8cm,height=5.5cm]{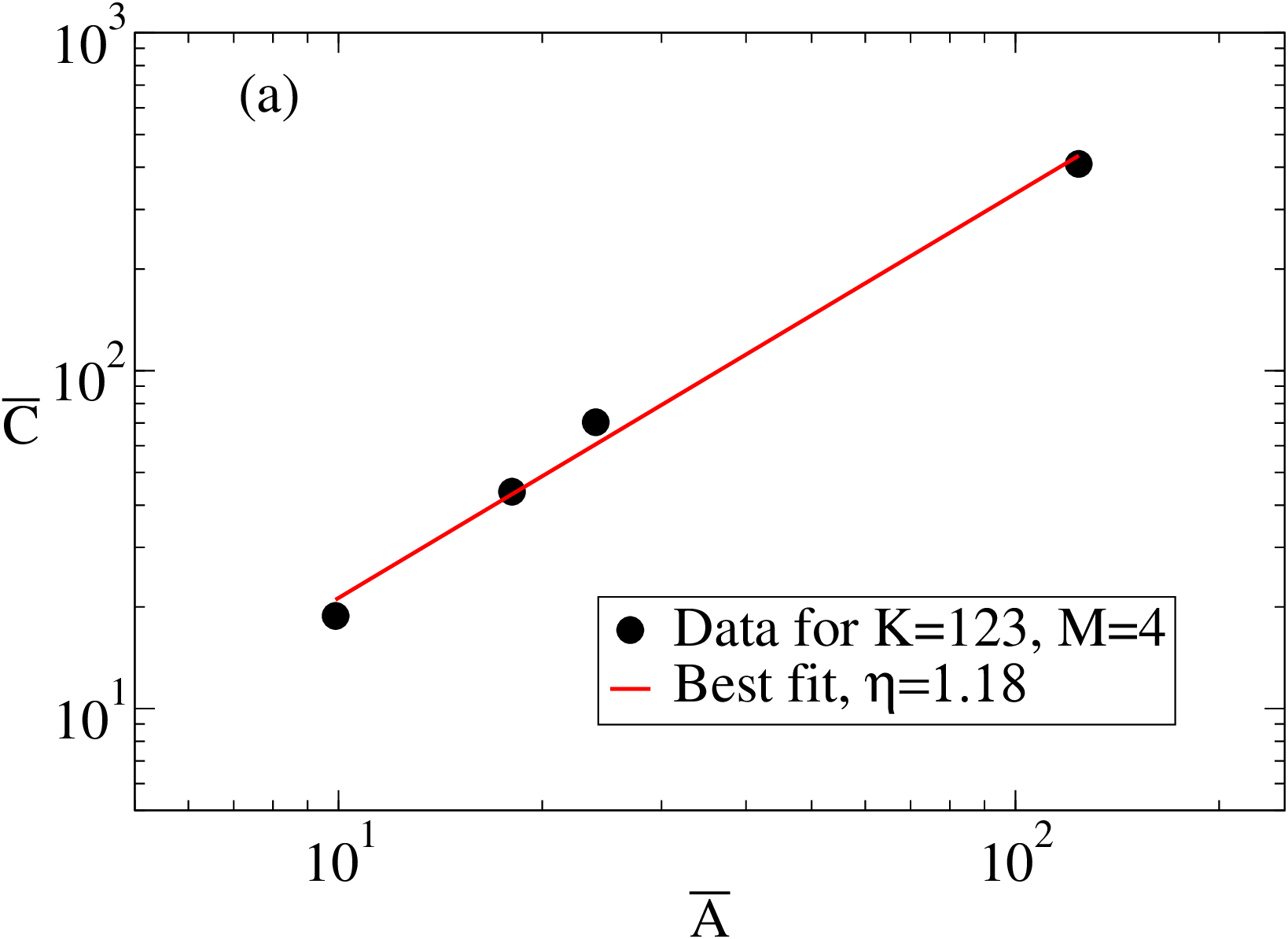}}
\vspace{0.4cm}
\centerline{\includegraphics[width=8cm,height=5.5cm]{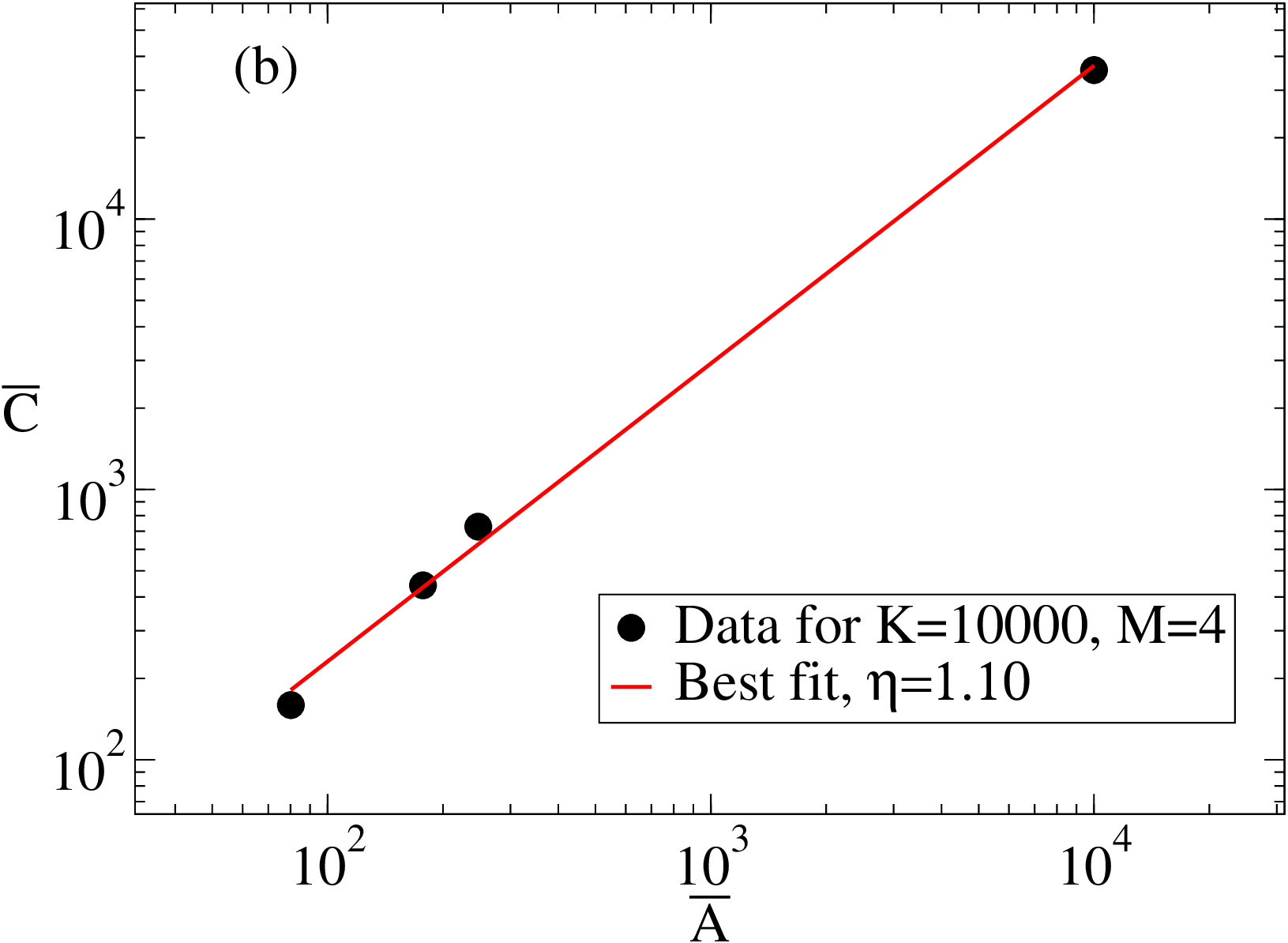}}
\caption{Log-log Plots for random networks. (a) $\overline{C}_\alpha\times\overline{A}_\alpha$ 
for a network with $K=123$ and $M=4$ (b)$\overline{C}_\alpha\times\overline{A}_\alpha$ graph 
for a random network with $K=10000$ and $M=4$. Note that when $K$ is large enough the exponent 
$\eta \approx 1$}
\label{fig1}
\end{figure}

In Fig.~\ref{fig1}(a) it is  shown the $\overline{C}_\alpha\times\overline{A}_\alpha$ 
graph for a random network with $K=123$, the same number of trophic specie that the 
Ythan Estuary with parasites, and $M=4$. A best fit furnishes $\eta=1.18$.
A similar fit for $K=93$, the same number of trophic specie that the 
Little Rock Lake food web, and $M=4$ give us $\eta=1.21$.
Note that the exponent decrease when $K$ increases. Clearly, our exponent is larger 
than that found by ~\cite{Garla1} for the same trophic species number $K$. 
But, when $K$ grows our exponent become smaller that them. Obviously, if
$\eta=1.13$ represent a universal value for food webs of all the sizes, then 
random spanning trees networks with the same number of trophic levels $M$ are 
more efficient than food webs. In fig.~\ref{fig1}(b) it is shown the 
$\overline{C}_\alpha\times\overline{A}_\alpha$ graph for a random network 
with $K=10000$ and $M=4$. Note that when $K$ is large enough the exponent $\eta \approx 1$

\begin{figure}[hd]
\centerline{\includegraphics[width=8cm,height=5.5cm]{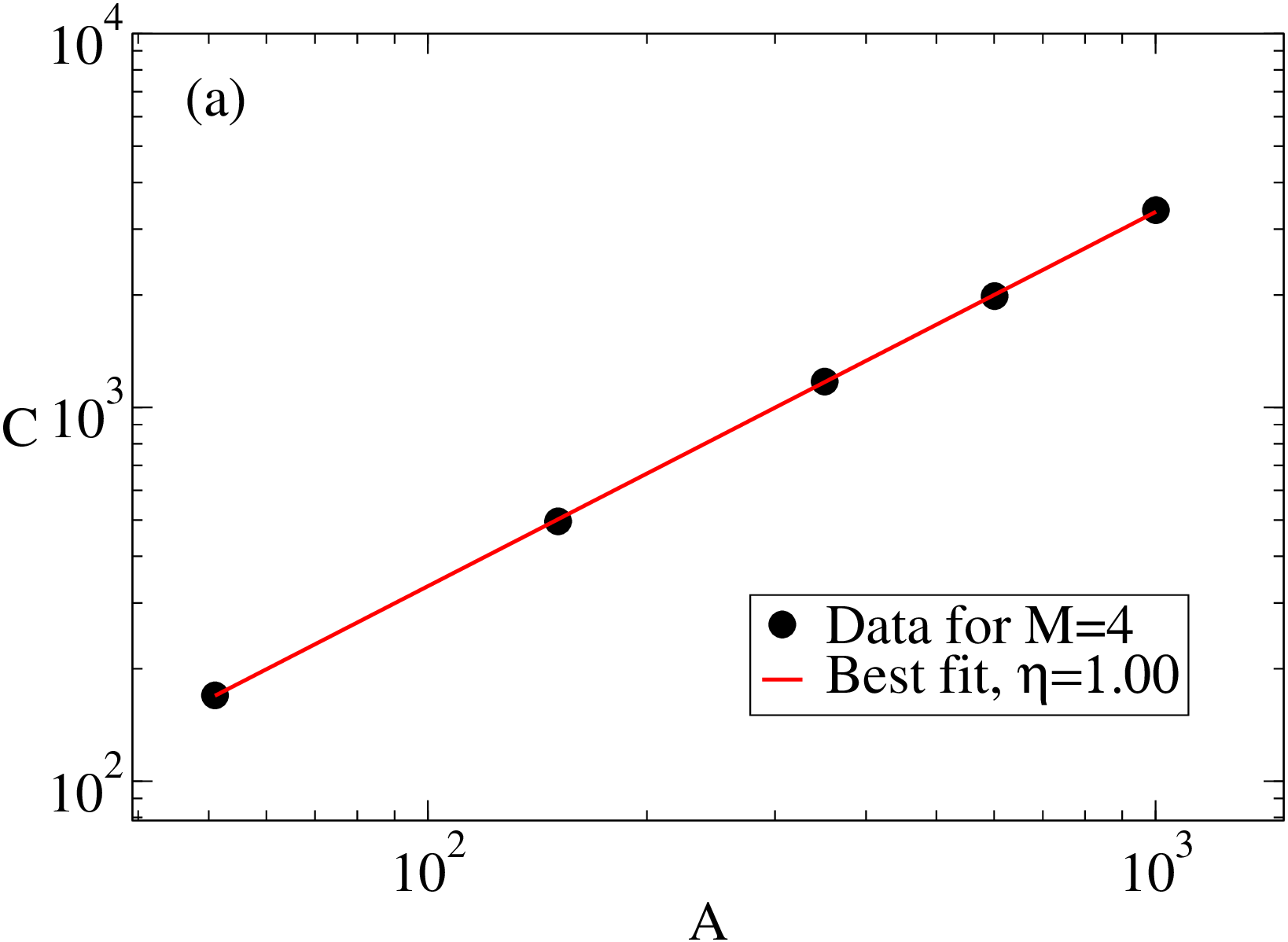}}
\vspace{0.4cm}
\centerline{\includegraphics[width=8cm,height=5.5cm]{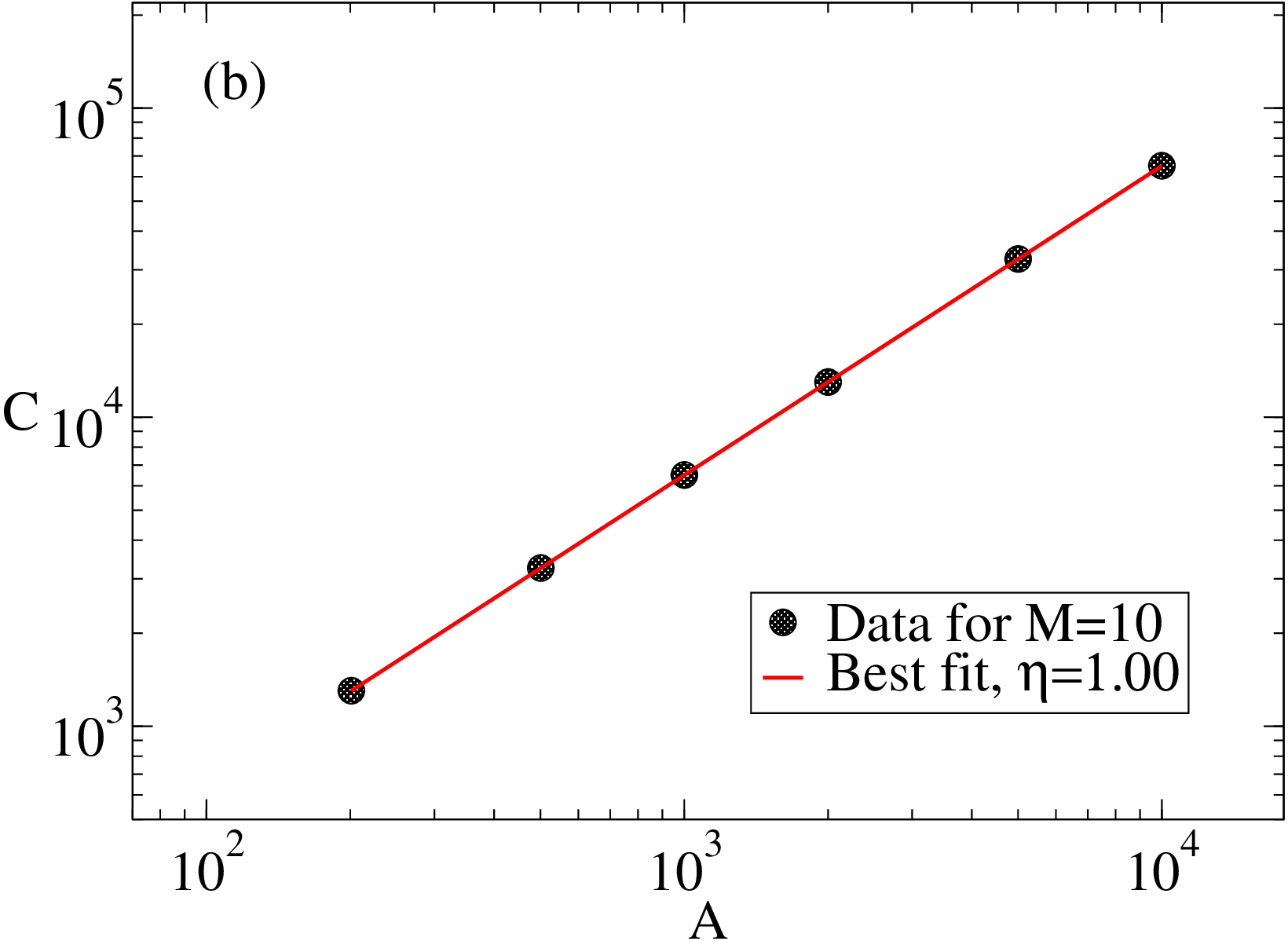}}
\caption{Log-log Plots for random networks. (a) $C_0\times A_0$ for $M=4$ and $K$ 
varying from $50$ up to $1000$ and (b) $C_0\times A_0$ for $M=10$ and $K$ varying from $200$ 
up to $5000$. Note that the large scale exponent has the same value 
$\eta= 1.00$ and is independent of $M$.}
\label{fig2}
\end{figure}

The exponent $\eta$ can also be computed by the procedure (b).
For each value of $K$ we perform an average for
several configurations and find the mean value of $C_0$. In Fig.~\ref{fig2}(a), it is shown
the $C_0\times A_0$ plot for random networks with $M=4$ and $K$ varying from $50$
up to $1000$. Now we have that $\eta=1.00$. It is worth
mentioning, that $C_0\times A_0$ always furnishes $\eta=1$
independently of the range of $K$. We have also simulated random networks
with $M=10$ trophic levels. In Fig.~\ref{fig2}(b) it is shown
the $C_0\times A_0$ plot. The results are similar.

Now let us present the central point of this paper, a general 
argument to demonstrate that the large scale exponent is $\eta =1$ for 
large $K$. Let us consider a spanning tree with $M$ and $K$, both fixed. 
To obey the constraint of $M$ fixed, we put one site in each level.
 Now we must put each one of the  reminder $K-M$ sites. Since $C_0$ is
 cumulative, a site put as near as possible of the environment has the
 minimal contribution to the global cost. On the other hand, a site put
 as far as possible of the environment has a maximal contribution to $C_0$.
  To construct the network with maximum value of $C_0$, $C_{0,max}$, we must
  link all $K-M$ sites to the site of the last level. In this network
  we have  $N_1=N_2=\dots=N_{M-1}=1$ and $N_M=K-M+1$. $C_{0,min}$ is
  obtained by linking the $K-M$ sites directly to the site representing the
  environment. In this case, we have that $N_1=K-M+1$ and 
 $N_2=N_3=\dots=N_M=1$. Note that these
constructions are the closest networks to the star-like and chain-like 
ones, respectively, that obey the constraints of $M$ and $K$ fixed.
 Using the Eq.~\ref{eq1} we have,

\begin{eqnarray*}
C_{0,min}&=&1+2K+\frac{M}{2}(M-1)\\
C_{0,max}&=&1+K(M+1)+\frac{M}{2}(1-M)
\end{eqnarray*}
Then, the lower and the upper bounds for the exponent $\eta$ are

\begin{eqnarray*}
\eta_{max}&=&\frac{\ln C_{0,max}}{\ln (K+1)}=
\frac{ \ln [ 1+K(M+1)+\frac{M}{2}(1-M) ]} {\ln (K+1)}~~,\\
\eta_{min}&=&\frac{\ln C_{0,min}}{\ln (K+1)}=
\frac{ \ln [ 1+2K+\frac{M}{2}(M-1)]} {\ln (K+1)}~~.
\end{eqnarray*}
When  $K\to\infty$, we have that $\eta_{max}=\eta_{min}\to 1$.

Consider again the simulation of random networks. We
verified that the constructions with minimum and maximum $C_0$
are the ones just described. Moreover, the result above explains why we
find $\eta\to 1$ when $K$ is large in the simulations of random networks.

In summary, we studied the transportation properties of several networks that
represent spanning trees of food webs. First, we analyzing an idealized
hierarchical model that can be analytically solved. Then we show that the exponent 
$\eta$ depends on value of $K$ and, in the limit that $K$ is large enough, the exponent
$\eta$ approaches the value $1$. After, we construct random networks 
that more realistically represents a spanning tree formed by food webs. We evaluate
numerically the exponent $\eta$ by several procedures. Again, in all cases the exponent
depends on size of web and if $K$ is large $\eta\to 1$. One important point is that all 
the results are independent of the number of trophic levels $M$. Moreover, 
 we establish a maximum
and a minimum values for the exponent $\eta$ in a general spanning tree with $K$ and $M$ fixed.
 When the number
of species is large these values  became equal to $1$. Therefore, we
 must find $\eta=1$ for a large food web and we can conclude that food webs are very
 efficient resource transportation systems.

The authors thank to Funda\c c\~ao de Amparo \`a Pesquisa do
Estado de Minas Gerais (FAPEMIG), Coordena\c c\~ao de Aperfei\c
coamento de Pessoal de N\'\i vel Superior (CAPES) and
to Conselho Nacional de Pesquisa (CNPq), Brazilian agencies.\\
$^{\ddagger}$  Electronic address: lbarbosa@fisica.ufmg.br\\
$^{\ast}$      Electronic address: alcides@fisica.ufmg.br\\
$^{\dagger}$   Electronic address: jaff@fisica.ufmg.br\\

\end{document}